\documentclass[11pt]{article}
\pdfoutput=1  
\usepackage{jheppub}

\usepackage{graphicx,amssymb,amsmath,color}
\usepackage{hyperref}

\def\beq{\begin{equation}}
\def\eeq{\end{equation}}

\def\bea{\begin{eqnarray}}
\def\eea{\end{eqnarray}}
\def\bei{\begin{itemize}}
\def\eei{\end{itemize}}
\def\bmat{\begin{matrix}}
\def\emat{\end{matrix}}
\def\={\,=\,}
\def\+{\,+\,}
\def\-{\,-\,}

\preprint{APCTP Pre2017-014}

\title{LSP baryogenesis and neutron-antineutron oscillations from R-parity violation}

\author[a]{Lorenzo Calibbi,}
\emailAdd{calibbi@itp.ac.cn}
\affiliation[a]{CAS Key Laboratory of Theoretical Physics, Institute of Theoretical Physics,
\\Chinese Academy of Sciences, Beijing 100190, P.~R.~China}

\author[b]{Eung Jin Chun,}
\emailAdd{ejchun@kias.re.kr}
\affiliation[b]{Korea Institute for Advanced Study, Seoul 02455, Korea}

\author[c,d]{Chang Sub Shin}
\emailAdd{changsub.shin@apctp.org}
\affiliation[c]{Asia Pacific Center for Theoretical Physics, Pohang 37673, Korea}
\affiliation[d]{Department of Physics, Postech, Pohang 37673, Korea}

\abstract{R-parity and baryon number violating operators can be allowed in the Supersymmetric Standard Model and thus lead to interesting baryon number violating processes such as $n$-$\overline n$ oscillations and baryogenesis of the Universe via the decay of the lightest supersymmetric particle (LSP). Adopting the LSP baryogenesis mechanism realized by the late decay of the axino, we identify a single coupling $\lambda''_{313}$ as a common origin for the matter-antimatter asymmetry of the Universe
as well as potentially observable $n$-$\overline n$ oscillation rates.
From this, rather strong constraints on the supersymmetry breaking masses and the axion decay constant are obtained.
The favoured parameter space of $\lambda''_{313} \sim 0.1$ and sub-TeV masses for the relevant sparticles is 
readily accessible by the current and future LHC searches.}

\begin{document}
\maketitle

\section{Introduction}
Unlike the Standard Model (SM), the Minimal Supersymmetric Standard Model (MSSM) may allow baryon and lepton number violating operators causing fast proton decay. Such a problem is often evaded by introducing a discrete symmetry, R-parity, enforcing baryon and/or lepton number conservation.  When both baryon and lepton number violation are forbidden, the lightest supersymmetric particle (LSP) becomes stable and thus can be a good Dark Matter (DM) candidate if it is neutral. This has been considered as one of the good motivations for supersymmetry. On the other hand, the proton stability can be guaranteed by imposing only lepton number conservation. In this case, among the possible R-parity violating (RPV) terms, only baryon number violating (renormalizable) terms of the form
\begin{equation} \label{WBNV}
W_{B \hspace{-0.5em} / \hspace{0.0em}}= {1\over2} \lambda''_{ijk} U^c_i D^c_j D^c_k
\end{equation}
are allowed in the MSSM superpotential. The above terms can lead to observable $\Delta B=2$ processes like neutron-antineutron oscillations \cite{nnbar-ILL,nnbar-SK,nnbar-ESS} and di-nucleon decays like $NN\to KK, \pi\pi$ \cite{NNKK,NNpipi} from a variety of diagrams involving supersymmetric particles and the $\lambda''_{ijk}$ interactions \cite{Z84, BM85, GS94, CK96,16calibbi}. A new experiment has been proposed at the European Spallation Source (ESS) 
with the aim of improving the sensitivity to the neutron-antineutron transition probability by 
up to three orders of magnitude \cite{nnbar-ESS}. 

The baryon number violation (BNV) allowed in the superpotential~(\ref{WBNV}) could be a source of  the matter-antimatter asymmetry of the universe \cite{sakharov91} through decays of the LSP that are induced by $\lambda''_{ijk}$ interactions. If baryogenesis occurred above the weak scale, 
a strong bound on $\lambda''_{ijk}$,
\begin{equation} \label{washout}
\lambda''_{ijk} \lesssim10^{-6},
\end{equation}
can be set by requiring not to wash out the baryon asymmetry for squark masses around the TeV scale \cite{CDEO91}.
Such small couplings are not sizable enough to generate the observed baryon asymmetry. 
Thus one has to rely on baryogenesis at a very low temperature.
Furthermore, the out-of-equilibrium decay of the LSP cannot lead to the desirable CP/baryon asymmetry at the usual second order of $\lambda''$ \cite{NW79}. For these reasons, almost all the existing models implement the BNV baryogenesis mechanism by late decays of supersymmetric particles other than the LSP  \cite{DH87, CR91, MR92, Cui13, Romp13, Covi15}.  
However, it was recognized in Ref.~\cite{MS15} that a  LSP baryogenesis can be realized  through the interference of a $\Delta B=1$ (four quark) and $\Delta B=2$ (six quark) operator at two loop.  In this scenario, the LSP is considered to be the axino, a supersymmetric partner of the axion, as its interactions are suppressed by an intermediate axion scale $f_a\approx 10^{10-12}$ GeV leading to the late decay required for baryogenesis.
Recall that the axion provides an elegant solution to the strong CP problem and, for values of the decay constant $f_a$ compatible with the above-mentioned range,  it is a good Dark Matter candidate \cite{jekim}.

\medskip

The purpose of this work is to investigate if the axino LSP baryogenesis can be implemented successfully by a certain BNV coupling which also leads to observable $n$--$\overline n$ oscillations. Following the idea of \cite{MS15}, we will consider the axino as the LSP decaying through a BNV coupling.
In Section \ref{sec:nnbar}, we will identify the BNV coupling $\lambda''_{313}$ as a promising candidate, which can lead to an observable $n$-$\overline n$ oscillation consistent with our baryogenesis mechanism.   
The axino lifetime strongly depends on how to realize the axion mechanism. The two typical models by DFSZ \cite{Dine:1981rt,Zhitnitsky:1980tq} and KSVZ \cite{Kim:1979if,Shifman:1979if} will be considered in Section \ref{sec:DFSZ} and \ref{sec:KSVZ}, respectively.  A discussion of the parameter space compatible with baryogenesis and large 
$n$-$\overline n$ oscillation rates as well as of the impact of LHC searches for RPV supersymmetry will be given in Section \ref{sec:LHC}.
We conclude in Section \ref{sec:conclusions}.
 
\section{Observable neutron-antineutron oscillations}
\label{sec:nnbar}
In the presence of a $\Delta B=1$ coupling in (\ref{WBNV}), the $\Delta B=2$ operator like $(udd)^2$ or $(uds)^2$ arises after integrating out heavy squark fields to induce $n$--$\overline n$ oscillations and/or di-nucleon decays.
Currently the most stringent (direct) bound on $\lambda''_{ijk}$ comes from the $NN\to KK$ search \cite{NNKK}:
$\lambda''_{112} < 10^{-6}-10^{-7}$
for the squark masses in the TeV region \cite{16calibbi}, which is comparable to (\ref{washout}). 
Varous couplings $\lambda''_{ijk}$ lead to the $n$--$\overline n$ oscillation operator
\begin{equation}
{\cal L}_{n \overline n} = C_{n\overline n} (udd)^2 + h.c.
\end{equation}
at tree or loop level in combination with flavor mixing among left-handed or right-handed squarks and possibly
left-right squark mixing \cite{Z84, BM85, GS94, CK96}. 
In fact, due to the contraction of the color indices in (\ref{WBNV}) through a totally antisymmetric tensor, 
the BNV couplings are antisymmetric under the exchange of the flavor indices of the $D^c$ superfields, $\lambda''_{ijk}=-\lambda''_{ikj}$, which implies that non-vanishing contributions to the $n$--$\overline n$ operator 
must involve squarks of the second or third generation mixing with the first generation.
As a consequence, some of the strongest bounds can be put only in combination with squark flavor mixing such as
$(\delta^d_{RR})_{ij}$ (which parameterizes the mixing among right-handed squarks): 
in \cite{16calibbi},  
again assuming supersymmetric masses around the TeV, it was found $\lambda''_{11k} (\delta^d_{RR})_{k1} \lesssim 10^{-8}$.  
\begin{figure}[t]
\begin{center}
\includegraphics[width=0.6\textwidth]{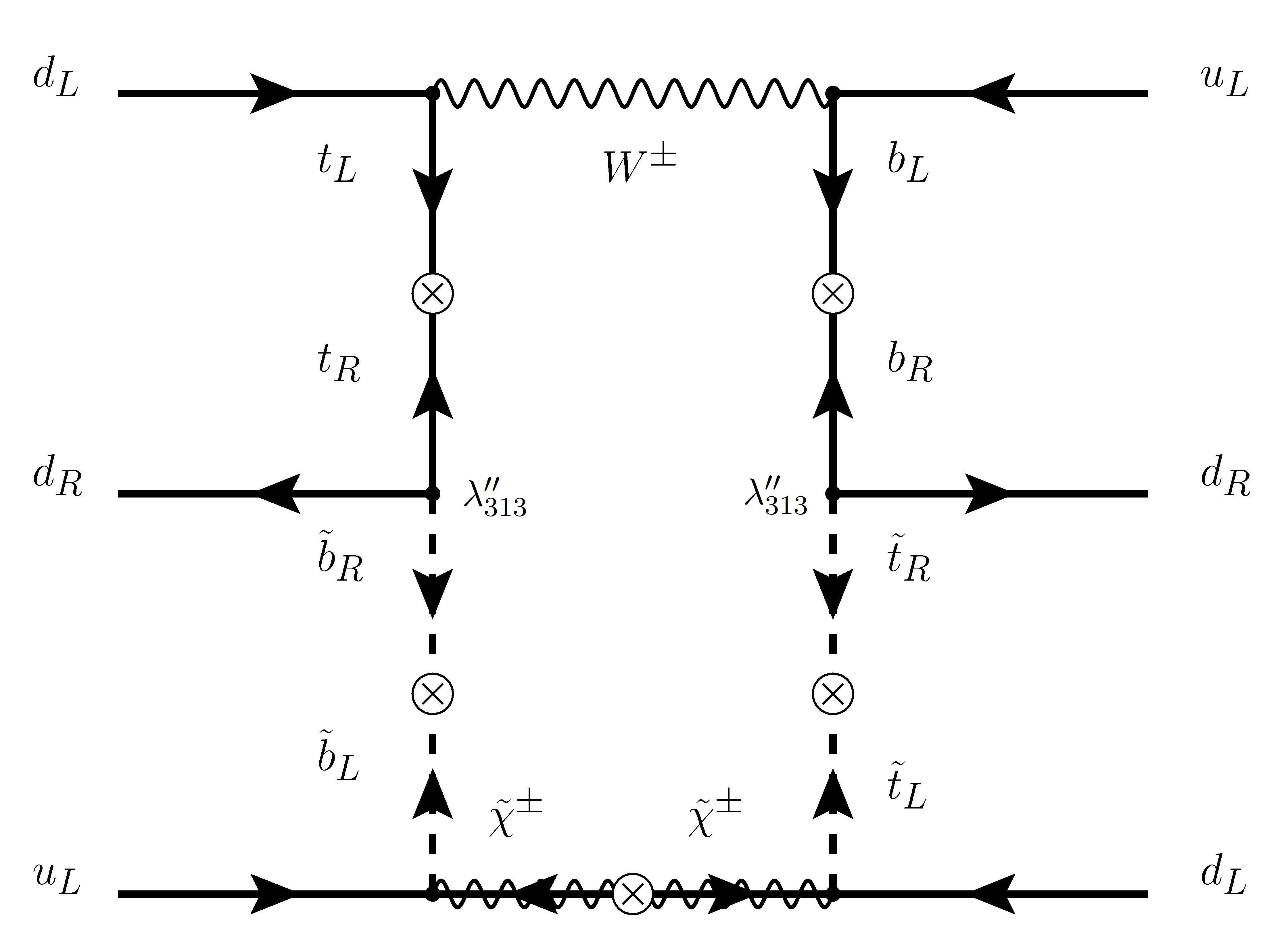}
\caption{Electroweak loop diagram inducing $n$--$\overline n$ oscillation from the $\lambda''_{313}$ coupling. \label{fig:CK}}
\end{center}
\end{figure}

Barring additional degrees of freedom, we assume that no squark flavor mixing arises from supersymmetry breaking. Then, the flavor mixing can only arise from electroweak loop corrections -- controlled by the Cabibbo-Kobayashi-Maskawa (CKM) matrix --
that, however, hardly induce a sizable $n$--$\overline n$ oscillation coefficient $C_{n \overline n}$. One can find among various contributions that observable $n$--$\overline n$ oscillations can be induced by 
$\lambda''_{113}$ or $\lambda''_{313}$ through electroweak one-loop diagrams \cite{GS94, CK96}. 
From the analysis of \cite{16calibbi}, one can see that the $\lambda''_{113}$ contribution to $C_{n\overline n}$ is much larger than the $\lambda''_{313}$ contribution (involving a suppression from smaller CKM entries) for the same set of parameters, and thus observable $n$--$\overline n$ rates from  $\lambda''_{113}$ requires smaller values of the BNV coupling or a rather heavier supersymmetric spectrum. Such values make the axino decay much later than the Bing-Bang Nucleosynthesis (BBN) epoch and thus can not lead to a viable baryogenesis as will be shown in the following section. Therefore, the $\lambda''_{313}$ contribution remains to be a more favorable option for observable $n$--$\overline n$ oscillations and baryogenesis. 

We update the original contribution proposed by Chang and Keung (CK)  \cite{CK96} (whose diagram is depicted in Fig.~\ref{fig:CK}) in a complete form properly taking the squark left-right mixing into account. Including only the  lightest squark (stop or sbottom) contribution we find
\begin{eqnarray}
C^{CK}_{n \overline n} && \!\! = {g^4\over 64 \pi^2} (\lambda''_{313})^2 (V_{td} V^*_{ub})^2 \,
m_{\widetilde \chi^\pm} m_t m_b c_{\widetilde t} s_{\widetilde t}  c_{\widetilde b} s_{\widetilde b} \\
&&~~~ J_6(m^2_{\widetilde t_1}, m^2_{\widetilde b_1}, m^2_{\widetilde \chi^\pm}, m^2_W, m^2_t, m^2_b)  
\nonumber\\
 \mbox{where}&&\!\!  J_6(a_1, a_2, a_3,a_4,a_5,a_6)=\sum_{i=1}^6 
{ a_i \log a_i \over \prod_{k\neq i} (a_k-a_i) }\, .\nonumber
\end{eqnarray}
Here  the effect of the left-right squark mixing is properly encoded in the squark mixing angle $\theta_{\widetilde q}$ through the combination of $c_{\widetilde q}s_{\widetilde q} \equiv  \cos\theta_{\widetilde q} \sin\theta_{\widetilde q}$.
Taking for simplicity $m_{\widetilde \chi^\pm}= m_{\widetilde t}=m_{\widetilde b} = m_S$  and  
maximal stop and sbottom mixing ($ 2 c_{\widetilde t} s_{\widetilde t}=1$ and  $ 2 c_{\widetilde b} s_{\widetilde b}=1$), 
one finds the $n$--$\overline n$ oscillation time $\tau_{n\overline n}=1/(C_{n\overline n}^{CK} \langle  n| (udd)^2 |\overline n\rangle)$ as follows:
\begin{equation}
\tau_{n\overline n} \approx 10^9 \sec \left(0.2 \over \lambda''_{313}\right)^2 \left( m_S \over 500 \mbox{ GeV}\right)^{5}  
\left( 0.5 \over c_{\widetilde t} s_{\widetilde t}\right)\left( 0.5 \over c_{\widetilde b} s_{\widetilde b}\right)
{(250 \mbox{ MeV})^6 \over \langle  n| (udd)^2 |\overline n\rangle} ,
\label{eq:nnbar-num}
\end{equation}
where we neglected and order-one prefactor variation in the loop function $J_6$ for different values of $m_S$.
Taking into account the large uncertainty (of one order of magnitude or more) in the hadronic matrix element 
$ \langle  n| (udd)^2 |\overline n\rangle$, the resulting oscillation time can be within the future sensitiviy limit of $\tau_{n \overline n} \simeq 3\times10^{9} \sec$ of the proposed experiment at the ESS \cite{nnbar-ESS}, if indeed a $\mathcal{O}(10^3)$ 
improvement on the limit set by \cite{nnbar-ILL} on the oscillation probability ($P_{n\overline n}\propto 1/\tau_{n\overline n}^2 $) is achieved. 
In terms of the limit on the oscillation time, the bound of \cite{nnbar-ILL} reads $\tau_{n \overline n} > 0.86\times10^{8} \sec$.
While this was obtained directly employing free neutrons, the indirect limit from bounded neutrons in Super-Kamiokande
is  $\tau_{n \overline n} > 2.7\times10^{8} \sec$ \cite{nnbar-SK}.

\section{DFSZ axino baryogenesis}
\label{sec:DFSZ}
\begin{figure}[t]
\begin{center}
\includegraphics[width=0.4\textwidth]{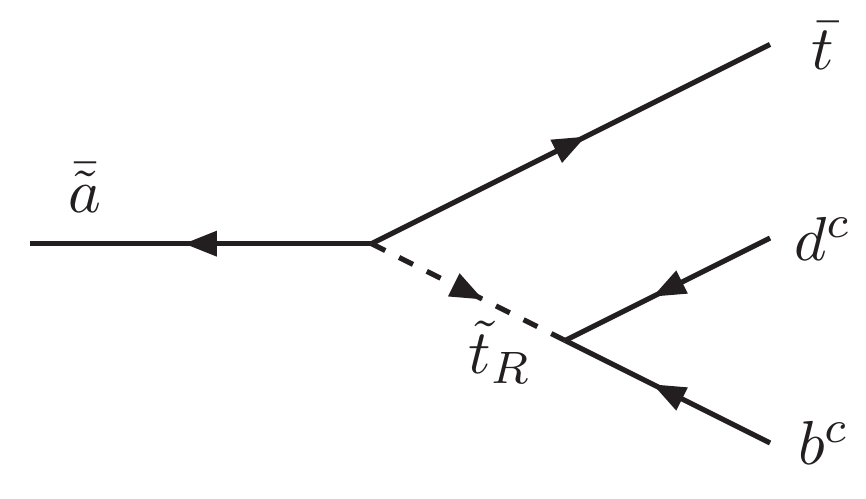} \
\includegraphics[width=0.4\textwidth]{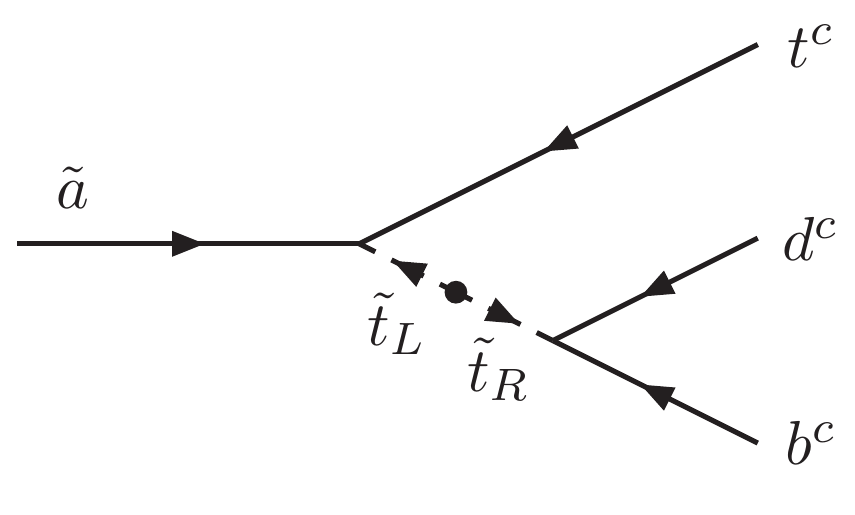} 
\caption{Tree-level diagrams for the decays  of DFSZ type axino, $\widetilde a$\label{fig:DFSZ_decay}}
\end{center}
\end{figure}

Let us first consider the DFSZ axion model to realize the axino LSP baryogenesis mechanism \cite{MS15}.
The axino ($\tilde a$) is the fermion component of the axion superfield, 
\begin{equation}
A = (s + i a)/\sqrt{2} + \sqrt{2}\theta \tilde a +\theta^2 F_A, 
\end{equation}
where $a$ is the axion, $s$ is the saxion field. 
The $U(1)_{\rm PQ}$ shift symmetry of the axion, under which $A\to A + i\alpha f_a$, is anomalously broken by the $SU(3)_C$ gauge symmetry. 
In the DFSZ axion model, the MSSM fields are also charged under $U(1)_{\rm PQ}$. So the relevant interactions between $A$ and the MSSM fields are given by the $\mu$-term superpotential: 
\begin{equation}
W = \mu e^{A/f_a} H_u H_d  = \mu H_u H_d +  \frac{\mu}{f_a} A H_u H_d +\cdots 
\end{equation}
From here we get the axino-Higgsino-Higgs interactions. 
Since the axino is the LSP in our scenario, the mixing between the axino and the Higgsino via the Higgs vacuum expectation value is important.  For the axino-quark-squark interactions induced by the axino-Higgsino mixing, 
the axino decay rate follows from diagrams as those in Fig.~\ref{fig:DFSZ_decay}.  The corresponding operator can be obtained after integrating out the heavy squarks:
\begin{eqnarray}
{\cal L}_{\rm decay}&=& {\lambda''_{ijk}\over f_a} \left(
 {m_{u_i} \over m_{\widetilde u_{i A}}^2} \, {\overline{\widetilde a}\overline u_i   d^c_j d^c_k }
+{e^{- i \varphi_{u_i}}m_{u_i}\over m_{\widetilde u_{i B}}^2}\,\widetilde a u_i d_j^c d_k^c   + h.c.\right)  \nonumber\\
&& + (u_i,\widetilde u_i \leftrightarrow d_j,\widetilde d_j) + (u_i,\widetilde u_i\leftrightarrow d_k,\widetilde d_k), 
\end{eqnarray}
where $\varphi_{u_i}\equiv {\rm Arg}(X_{u_i})$, with 
$X_{u_i} = A_{u_i} - \mu^* \cot\beta$ being the parameter that controls the squark left-right mixing, and 
\begin{eqnarray}
{1 \over m_{\widetilde u_{i A}}^2} \equiv{ \cos^2\theta_{\widetilde u_i} \over  m^2_{\widetilde u_{i1}}} +{\sin^2\theta_{\widetilde u_i}\over m^2_{\widetilde u_{i2}}},\  \  
{1\over m_{\widetilde u_{i B}}^2}\equiv { \cos\theta_{\widetilde u_i}\sin\theta_{\widetilde u_i}\over m_{\widetilde u_{i1}}^2} 
-{\cos\theta_{\widetilde u_i}\sin\theta_{\widetilde u_i}\over m_{\widetilde u_{i 2}}^2 },
\end{eqnarray}
for the up-type left-right squark mixing angle $\theta_{\widetilde u_i}$, and the corresponding squark mass eigenvalues, $m_{\widetilde u_{i 1}}$ and $m_{\widetilde u_{i 2}}(> m_{\widetilde u_{i1}})$.
Considering the coupling $\lambda''_{313}$, the axino decay is dominated by the top-quark channel mediated by $m_{\widetilde t_1}$:
\begin{equation} \label{Ldecay}
{\cal L}_{\rm decay} \simeq { \lambda''_{313}  
m_{t} \over f_a m_{\widetilde t_1}^2} \Big(
   c^2_{\widetilde t}   \, 
{\overline{\widetilde a} \overline t  d^c  b^c }  +
 c_{\widetilde t} s_{\widetilde t} e^{- i  \varphi_{\widetilde t}}   \, 
{\widetilde a t^c d^c b^c }  \Big)+ h.c.
\end{equation}
where we neglected the heavier stop contribution and defined $\varphi_{\widetilde t}\equiv \mbox{Arg}(X_t)$.
For a successful late baryogenesis, the decay temperature $T_D$ of the axino LSP should be much smaller than  the  supersymmetry breaking scale but larger than the BBN temperature, that is, $1 \mbox{ MeV} 
\lesssim T_D \ll m_{\rm SUSY}$. 
The decay rate reads 
\begin{equation}
\Gamma_{\widetilde a} \simeq { |\lambda''_{313}|^2 \over 256\pi^3}  
{m^2_{t}  |m_{\widetilde a}|^5\over m^4_{  \widetilde t_1} f^2_a},
\end{equation} and from this
we get for the axino decay temperature $T_D \approx \sqrt{ \Gamma_{\widetilde a} m_P}$:
\begin{equation}
T_D \approx 800 \mbox{ MeV} 
 \left( |\lambda''_{313}| \over 0.2 \right)
\left( 500 \mbox{ GeV} \over m_{\widetilde t_1}\right)^2
\left( |m_{\widetilde a}| \over 400 \mbox{ GeV} \right)^{5/2}
\left( 10^{10} \mbox{GeV} \over f_a \right),
\label{eq:decay_DFSZ}
\end{equation}
which can be easily consistent with the BBN bound of $T_D\gtrsim 1$ MeV. 
Notice that the other couplings $\lambda''_{i jk}$ with $i \neq 3$ can hardly satisfy the BBN bound due to the quark mass suppression of the axino coupling $ \propto m_q/f_a$. For $\lambda''_{313}={\cal O}(0.1)$, the NLSP will always prefer to decay into the SM particles, much faster than the BBN time. 
Therefore it is quite safe from the cosmological constraints. Instead it could give the interesting collider phenomenology which will be discussed in Section \ref{sec:LHC}.

\begin{figure}[t]
\begin{center}
\includegraphics[width=0.4\textwidth]{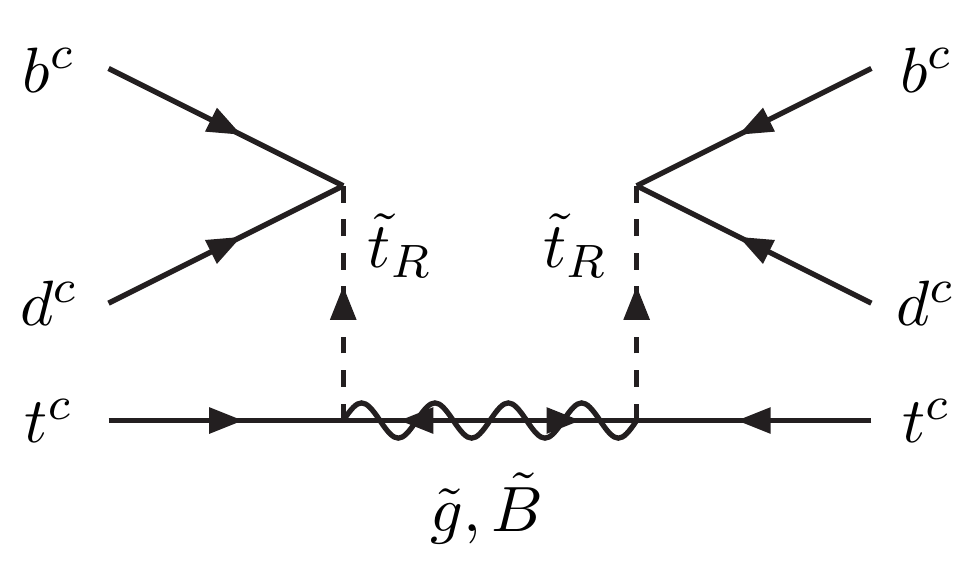} \
\includegraphics[width=0.4\textwidth]{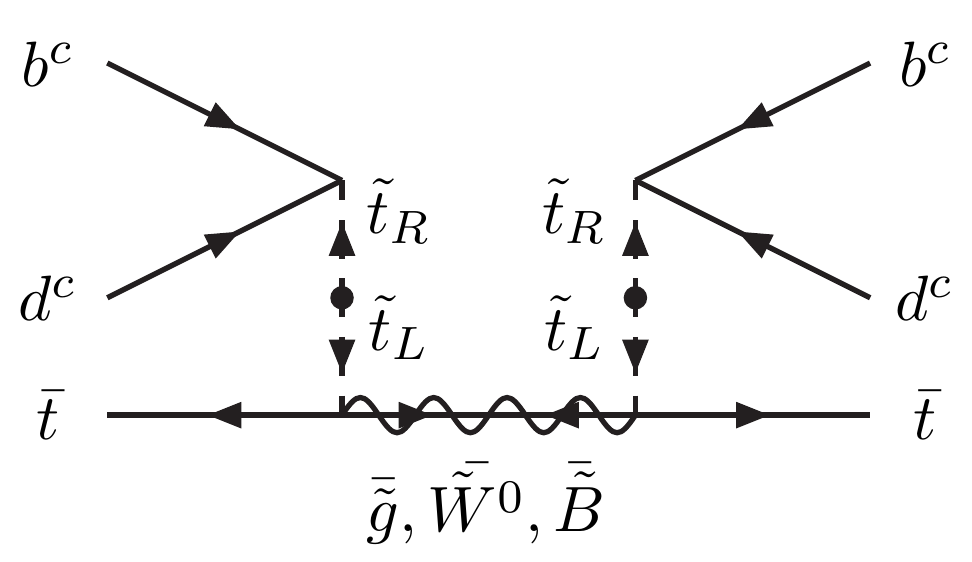} \\
\includegraphics[width=0.4\textwidth]{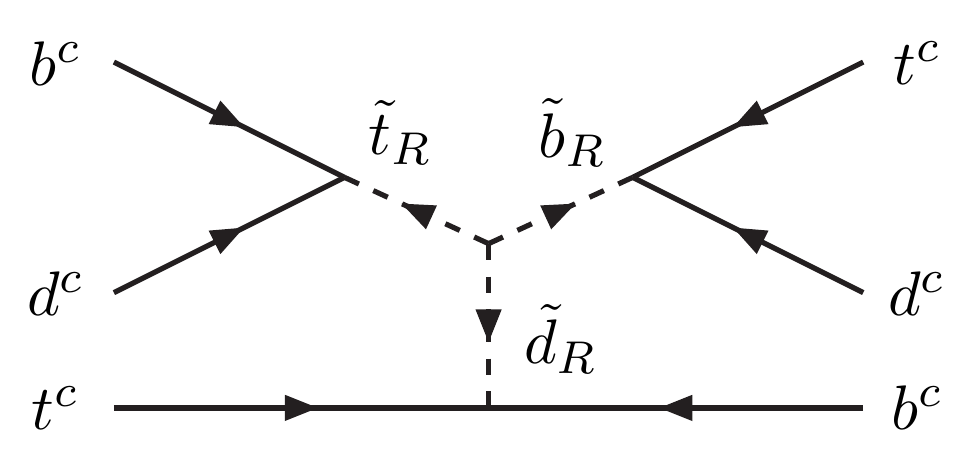} 
\caption{The six quark baryon number violating interactions mediated by gaugino and stop (top), and $A$-term and three squarks (bottom).\label{fig:Sixquark}}
\end{center}
\end{figure}

A CP asymmetry in the axino decay, as customary defined as
\begin{equation}
\epsilon\equiv \frac{\Gamma(\widetilde a\to X)- \Gamma(\widetilde a\to \overline{X})}{\Gamma(\widetilde a\to X)+ \Gamma(\widetilde a\to \overline{X})},
\end{equation}
is generated by the interference between the tree-level diagrams in Fig.~\ref{fig:DFSZ_decay} and the two-loop diagrams obtained by joining the $\Delta B =1$ diagrams of Fig.~\ref{fig:DFSZ_decay} with the $\Delta B =2$ ones shown in Fig.~\ref{fig:Sixquark}.
The calculation of the asymmetry gets simplified by considering the 6-quarks  $\Delta B =2$ operators that the diagrams 
of Fig.~\ref{fig:Sixquark} give rise to, once the supersymmetric particles are integrated out.
In general, the $\Delta B=2$ operator $(tdb)^2$ is dominantly generated through the stop-stop-gluino exchange:
\begin{eqnarray} \label{LB2}
{\cal L}_{\Delta B=2} &=&  
  { g_s^2 (\lambda''_{313})^2     \over  3  |m_{\widetilde g}|
\widetilde m^4_{\widetilde t_1} }  
\left(c^4_{\widetilde t} e^{-i \varphi_{\widetilde g}}\,  (t^c d^c b^c)^2 +
   c^2_{\widetilde t}s_{\widetilde t}^2e^{-i (2\varphi_{\widetilde t} - \varphi_{\widetilde g})} \,  (\overline t d^c b^c)^2\right)  +   h.c., 
\end{eqnarray}
where $\varphi_{\widetilde g} = \mbox{Arg}(m_{\widetilde g})$. 
There are also contributions from the stop-sbottom-gluino exchange for $m_{\widetilde b_1}\simeq m_{\widetilde t_1}$.  
When the gluino is relatively heavier than the Wino-like neutralino $\widetilde W^0$ 
(i.e.~if $m_{\widetilde g} \gtrsim (g_3/g)^2 m_{\widetilde W}\simeq 3 m_{\widetilde W}$), 
the contribution from the stop-stop-$\widetilde W^0$ diagram is also important:
\begin{eqnarray} \label{LB2_2}
{\cal L}_{\Delta B=2} &=&  
 { g^2 (\lambda''_{313})^2     \over   4|m_{\widetilde W}|
\widetilde m^4_{\widetilde t_1^c}   }  
\left(   c^2_{\widetilde t}s_{\widetilde t}^2e^{-i (2\varphi_{\widetilde t} - \varphi_{\widetilde W})} \,  (\overline t d^c b^c)^2\right)  +   h.c..
\end{eqnarray}
Finally, the contribution from the RPV trilinear soft mass, $A''_{313} = |A''_{313}|e^{ i\varphi_{313}}$, reads
\beq\label{LB2_3}
{\cal L}_{\Delta B= 2} =  \frac{|A''_{313} (\lambda''_{313})^2| (\lambda''_{313})^2 }{m_{\widetilde t_1}^2 m_{\widetilde d_1}^2 m_{\widetilde b_1}^2} \left(c_{\widetilde t}^2 c_{\widetilde d}^2 c_{\widetilde b}^2 e^{- i\varphi_{313}} (t^cd^c b^c)^2 \right).
\eeq

From (\ref{Ldecay}, \ref{LB2}, \ref{LB2_2}, \ref{LB2_3}), we find the CP asymmetry from the axino decay:
\begin{eqnarray}
\label{eq:asymmetry_DFSZ}
\epsilon &=&
\left|{ c_{\widetilde t}^2(c_{\widetilde t}^2-s_{\widetilde t}^2)g_s^2(\lambda''_{313})^2m_{\widetilde a}^5\over 32\pi^3  m_{\widetilde g} m_{\widetilde t_1}^4 }\right|  
{\rm Im} \left[ \frac{   m_t^2}{|m_{\widetilde a}|^2}  e^{i(\varphi_{\widetilde g}+\varphi_{\widetilde a})}  
 + \frac{ c_{\widetilde t}s_{\widetilde t}  m_t}{2|m_{\widetilde a}|} e^{i(\varphi_{\widetilde g} - \varphi_{\widetilde t})}\right] \\
  &+&\left|{ 3 c_{\widetilde t}^2s_{\widetilde t}^2 g^2 (\lambda''_{313})^2m_{\widetilde a}^5 \over 128\pi^3  m_{\widetilde t_1}^4  m_{\widetilde W}}\right|  {\rm Im}
 \left[ \frac{  s_{\widetilde t}^2 m_t^2}{|m_{\widetilde a}|^2}  e^{-i(\varphi_{\widetilde W}+\varphi_{\widetilde a})} 
 + \frac{c_{\widetilde t}  s_{\widetilde t} m_t}{2|m_{\widetilde a}|} e^{-i(\varphi_{\widetilde W} - \varphi_{\widetilde t})}
 + \frac{c_{\widetilde t}^2}{4}  e^{ i (2\varphi_{\widetilde t} - \varphi_{\widetilde W} + \varphi_{\widetilde a})} \right] 
 \nonumber\\
 &+&\left|{ c_{\widetilde t}^2 c_{\widetilde d}^2 c_{\widetilde b}^2(\lambda''_{313})^4A''_{313}m_{\widetilde a}^5\over 32\pi^3  m_{\widetilde t_1}^2 m_{\widetilde b_1}^2 m_{\widetilde d_1}^2 } \right|
 {\rm Im}\left[ \frac{ c_{\widetilde t}^2 m_t^2}{|m_{\widetilde a}|^2}   e^{i(\varphi_{313}+\varphi_{\widetilde a})}   
 + \frac{c_{\widetilde t}s_{\widetilde t}m_t}{2|m_{\widetilde a}|} e^{i(\varphi_{\widetilde 313} - \varphi_{\widetilde t})}  
 + \frac{s_{\widetilde t}^2}{4}  e^{ -i (2\varphi_{\widetilde t} - \varphi_{\widetilde 313} + \varphi_{\widetilde a})} \right]  \nonumber
\end{eqnarray} 
where $\varphi_{\widetilde a}$ is the phase of the axino mass, $m_{\widetilde a}$. 
For the gluino contribution,  we see some additional suppressions compared to $\widetilde W^0$ and $A''_{313}$ term contributions because the strong interaction does not distinguish $q$ and $q^c$, that is, gluino-quark-squark interactions preserve the charge conjugation symmetry while the weak interaction and RPV terms strongly violate it. In particular, unlike the others, the gluino contribution is always proportional to powers of $m_t/m_{\widetilde a}$ and vanishes for maximal left-right stop mixing: this reflects what we have just mentioned, namely that an asymmetry arises only in presence of chirality breaking.
When the gauginos ($\widetilde \lambda =\widetilde g,\, \widetilde W^0$) are light so that $m_{\widetilde \lambda} \sim m_{\widetilde a}$, the gaugino mass in the numerator should be substituted by $1/m_{\widetilde \lambda}\to m_{\widetilde \lambda}^*/(m_{\widetilde \lambda}^2 - m_{\widetilde a}^2)$, which can provide a resonant enhancement to the asymmetry.  Such an approximation is valid as long as $m_{\tilde \lambda} - m_{\tilde a} \gg \Gamma_{\tilde \lambda}$, a condition
which we assume hereafter. 

The CP asymmetry displayed in Eq.~(\ref{eq:asymmetry_DFSZ}) depends on a number of unknown phases, some of which need to be $\mathcal{O}(1)$ in order to trigger a successful baryogenesis, as we are going to see. On the other hand, for TeV-scale supersymmetric masses, large phases in the sfermion and gaugino sectors would be tightly constrained by the experimental bounds on electric dipole moments (e.g.~of the neutron and the electron), unless certain relations among generally independent phases are assumed (for a review see \cite{pospelov}). For simplicity here we are going to assume that the only large phase is the axino mass one, $\varphi_{\widetilde a}$, which is left unconstrained by low-energy observables. 

Numerically, the expression in Eq.~(\ref{eq:asymmetry_DFSZ}) gives $\epsilon \lesssim {\cal O}(10^{-7}- 10^{-6})$
for a choice of the parameters in the ballpark of Eq.~(\ref{eq:nnbar-num}), which give  potentially large $n-\overline{n}$ oscillation rates.
The interplay between baryogenesis and  $n-\overline{n}$ oscillation will be discussed in greater detail in Section \ref{sec:LHC} together with the impact of searches for supersymmetric partners at the LHC.
Provided that the stop mixing is large but not maximal, the gluino and wino contributions, i.e.~the first and second lines of Eq.~(\ref{eq:asymmetry_DFSZ}), give comparable contributions, while the A-term contribution (third line) is subdominant for $\lambda''_{313}=\mathcal{O}(0.1)$, as it is comparatively suppressed by a factor $(\lambda''_{313})^2$.

In order to achieve the observed baryon asymmetry, $Y_{\Delta B} \simeq 0.8\times10^{-10}$ \cite{planck}, a value of the CP asymmetry around $\epsilon={\cal O}(10^{-7})$ requires for the initial axino abundance $Y_{\widetilde a} = n_{\widetilde a}/s \approx 10^{-3}$, which could arise from the thermal production of the DFSZ axino, $Y_{\widetilde a}^{TP}$ for the reheating temperature greater than the Higgsino mass, and $f_a\lesssim 10^{10}\,{\rm GeV}$ \cite{Yaxino}. 
The actual value of $Y_{\widetilde a}$ can be depleted from the initial (thermal) abundance because the long-lived axino can dominate the energy density of the Universe. Depending on the decay temperature, the final yield value is  
 \beq
 Y_{\widetilde a}  = {\rm min}\left[ Y_{\widetilde a}^{TP},  \frac{3}{4}\frac{T_D}{m_{\widetilde a}}\right].
 \eeq
Therefore  the following constraint on the decay temperature (\ref{eq:decay_DFSZ}) is imposed by a successful baryogenesis: 
\beq
T_D \gtrsim \left(\frac{m_{\widetilde a}}{\rm TeV}\right)\left(\frac{\epsilon}{10^{-7}}\right)\,{\rm GeV}
\eeq
which
is stronger than that from BBN.
As we can see, this bound is fulfilled for values of the parameters in Eq.~(\ref{eq:decay_DFSZ}), translating in particular on a limit on the axion scale, $f_a\lesssim 10^{10}$ GeV. For these values of $f_a$, the axion misalignment mechanism can give a sizable contribution (up to 100\%) to the observed DM abundance, provided a rather large value of the pre-inflation misalignment angle \cite{nature}.

\section{KSVZ axino baryogenesis}
\label{sec:KSVZ}
\begin{figure}[t]
\begin{center}
\includegraphics[width=0.4\textwidth]{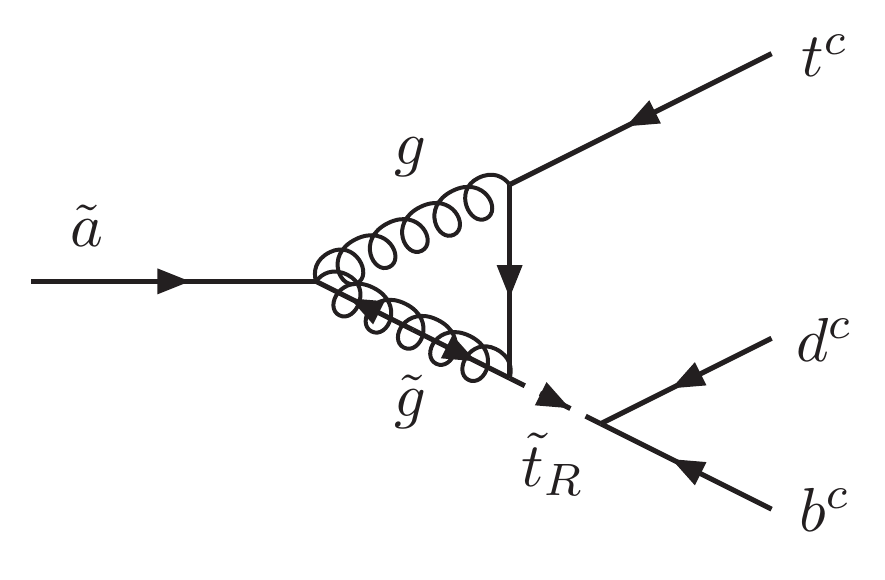} \
\includegraphics[width=0.4\textwidth]{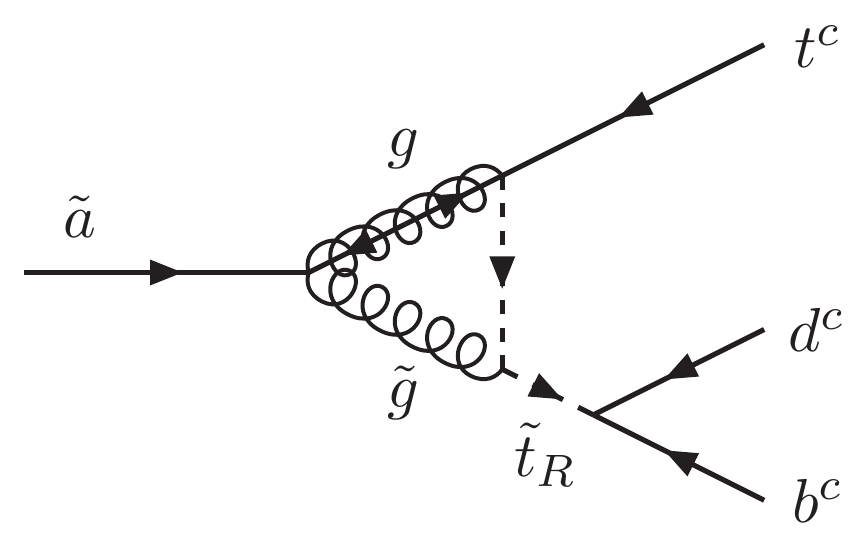} 
\caption{The 1-loop diagrams for the decays  of KSVZ type axino, $\widetilde a$\label{fig:KSVZ_decay}}
\end{center}
\end{figure}
In the KSVZ axion model, the MSSM particles are neutral under the $U(1)_{PQ}$ symmetry, so there is no
tree-level axino-squark-quark coupling. 
The leading interaction between the axino and the MSSM fields are given by the the anomalous couplings induced by  
\begin{equation}
{\cal L}\supset \int d\theta \frac{c_a}{16\pi^2 f_a} A {\cal W}^{a\alpha} {\cal W}^a_\alpha, 
\end{equation}
where ${\cal W}^a$ is the field strength chiral superfield for $SU(3)_C$. The axino can decay to three quarks at one loop through this axino-gluino-gluon interaction as shown in Fig.~\ref{fig:KSVZ_decay}.  After integrating out the squarks, we get the following effective Lagrangian for the axino decay:
\begin{eqnarray}
{\cal L}_{\rm decay} 
&=&  \frac{ g_s^4 }{(16\pi^2)^2} \frac{ \lambda''_{ijk}}{f_a} \ln\frac{f_a^2}{|m_{\widetilde g}^2|} 
\left(\frac{e^{- i\varphi_{\widetilde g}}|m_{\widetilde g}|}{m_{\widetilde u_{i A}}^2}\, \widetilde a u_i^c  d_j^c d_k^c  
+ { e^{-i (\varphi_{\widetilde u_i} - \varphi_{\widetilde g})} |m_{\widetilde g}|  \over m_{\widetilde u_{iB}}^2 }\, 
\overline{\widetilde a}\overline u_i^c d_j^c d_k^c + h.c.  \right)   \nonumber\\
&& + (u_i,\widetilde u_i \leftrightarrow d_j,\widetilde d_j) + (u_i,\widetilde u_i\leftrightarrow d_k,\widetilde d_k)
\end{eqnarray}
As in the DFSZ case, the dominant Lagrangian term for the axino decay is 
\begin{equation} \label{Ldecay_KSVZ}
{\cal L}_{\rm decay} \simeq \frac{g_2^4}{(16\pi^2)^2}{ \lambda''_{313}  
|m_{\widetilde g}| \over f_a m_{\widetilde t_1}^2} \ln\frac{f_a^2}{|m_{\widetilde g}|^2}\Big(
   c^2_{\widetilde t} e^{-i \varphi_{\widetilde g}}   \, 
{\widetilde a t^c d^c b^c }   +
 c_{\widetilde t} s_{\widetilde t} e^{- i  (\varphi_{\widetilde t} -\varphi_{\widetilde g})}   \, 
 {\overline{\widetilde a} \overline t  d^c  b^c }   \Big)+ h.c.
\end{equation}
The corresponding axino decay temperature is 
\beq
\label{eq:TD_KSVZ}
T_D \simeq 200\, {\rm MeV}\left(\frac{|\lambda''_{313}|}{0.2}\right) 
\left(\frac{500\,{\rm GeV}}{m_{\widetilde t_1}}\right)^2 
\left(\frac{|m_{\widetilde a}|}{400\,{\rm GeV}}\right)^{5/2} 
\left(\frac{|m_{\widetilde g}|}{2\,{\rm TeV}}\right)\left(\frac{10^{9}\, {\rm GeV}}{f_a}\right).
\eeq
Compared to the DFSZ case, we obviously need smaller $f_a$ to get a sizable decay temperature. 
The axino decay will generate the baryon asymmetry by the same ${\cal L}_{\Delta B=2}$ operators in Eqs.~(\ref{LB2}, \ref{LB2_2}). 
The asymmetry parameter is rather insensitive to the decay rate, and thus we get a similar result as in the DFSZ case:
\begin{eqnarray}
\epsilon &=&
\left|{ c_{\widetilde t}^2(c_{\widetilde t}^2-s_{\widetilde t}^2)g_s^2(\lambda''_{313})^2
m_{\widetilde a}^5\over 32\pi^3  m_{\widetilde g} m_{\widetilde t_1}^4 }\right|  
{\rm Im} \left[  \frac{ c_{\widetilde t}s_{\widetilde t}  m_t}{2|m_{\widetilde a}|} e^{i(\varphi_{\widetilde g} - \varphi_{\widetilde t})}  
 + \frac{1}{4}  e^{-i(\varphi_{\widetilde g} +\varphi_{\widetilde a})}  \right]  \\
  &+&\left|{ 3 c_{\widetilde t}^2s_{\widetilde t}^2 g^2 (\lambda''_{313})^2m_{\widetilde a}^5 \over 128\pi^3  m_{\widetilde t_1}^4  m_{\widetilde W}}\right|  
{\rm Im} \left[ \frac{  c_{\widetilde t}^2 m_t^2}{|m_{\widetilde a}|^2}  e^{-i( 2\varphi_{\widetilde g}+\varphi_{\widetilde W} - 2\varphi_{\widetilde t}+\varphi_{\widetilde a})}  
 + \frac{c_{\widetilde t}  s_{\widetilde t} m_t}{2|m_{\widetilde a}|} e^{-i(\varphi_{\widetilde W} - \varphi_{\widetilde t})}  
 + \frac{s_{\widetilde t}^2}{4} e^{ i (2\varphi_{ \widetilde g} - \varphi_{\widetilde W} + \varphi_{\widetilde a})}  \right]\nonumber\\
 &+&\left|{ c_{\widetilde t}^2 c_{\widetilde d}^2 c_{\widetilde b}^2(\lambda''_{313})^4A''_{313}m_{\widetilde a}^5\over 32\pi^3  m_{\widetilde t_1}^2 m_{\widetilde b_1}^2 m_{\widetilde d_1}^2 } \right|
{\rm Im} \left[ \frac{ s_{\widetilde t}^2 m_t^2}{|m_{\widetilde a}|^2}  e^{i(2\varphi_{\widetilde g} +\varphi_{313}-2\varphi_{\widetilde t}+\varphi_{\widetilde a})}  
 + \frac{c_{\widetilde t}s_{\widetilde t}m_t}{2|m_{\widetilde a}|} e^{i(\varphi_{ 313} - \varphi_{\widetilde t})}
 + \frac{c_{\widetilde t}^2}{4} e^{ -i (2\varphi_{\widetilde g} - \varphi_{  313} + \varphi_{\widetilde a})} \right]. \nonumber 
\end{eqnarray}  
Again, one has to take the replacement: $1/m_{\widetilde \lambda}\to m_{\widetilde\lambda}^*/(m_{\widetilde \lambda}^2- m_{\widetilde a}^2)$ 
for the gauginos $\widetilde\lambda=(\widetilde g,\, \widetilde W^0)$ when their masses are close to the axino mass $m_{\widetilde a}$. 
Comparing the above expression with Eq.~(\ref{eq:asymmetry_DFSZ})  for the DFSZ case, we see that  the gluino contribution has a term which is not suppressed by 
$m_t/m_{\widetilde a}$, because   in the KSVZ case the axino decay is mediated by the gluino-quark-squark interaction which does not flip the chirality, while in the DFSZ case the axino decay rate is mediated by the Higgsino-quark-squark interactions which flip the chirality.
This can make the asymmetry somewhat larger but still of the same order of magnitude,
$\epsilon \lesssim {\cal O}(10^{-7}- 10^{-6})$.

The KSVZ axino thermal production is more active at higher temperature as long as $T <f_a$, so that the final yield is sensitive to the reheating temperature of the Universe. Numerically,  $Y_{\widetilde a}^{TP}\propto T_{\rm reh}$.  For a sufficiently high $T_{\rm reh}$, a sizable amount of $Y_{\widetilde a}^{TP}$ can be easily obtained. For example, when $f_a\sim 10^9\,{\rm GeV}$, $T_{\rm reh} \gtrsim 10^5\,{\rm GeV} $ 
is enough to make $Y_{\tilde a}^{\rm TP}\sim 10^{-3}$ \cite{Yaxino_KSVZ}. 
Furthermore, such a thermal abundance can be reached even up to $f_a \sim 10^{11}$ GeV if
the heavy quark mass is considerable smaller than the axion scale \cite{Bae:2011jb}.
Including the case of axino dominated Universe before it decays, the actual yield value is
 \beq
 Y_{\widetilde a}  = {\rm min}\left[ Y_{\widetilde a}^{TP},  \frac{3}{4}\frac{T_D}{m_{\widetilde a}}\right] \gtrsim 
 10^{-3}\left(\frac{\epsilon}{10^{-7}}\right)^{-1}. 
 \eeq
From the expression for $T_D$, Eq.~(\ref{eq:TD_KSVZ}), we find that the above bound can be fulfilled for values of $f_a$ comparatively lower than in the DFSZ case. This makes it more unlikely to account for the full observed 
DM abundance in the KSVZ case.

\section{Discussion}
\label{sec:LHC}
In this section we give a more quantitative discussion of the interplay between LSP baryogenesis and $n-\overline{n}$ oscillations. As we have seen in the previous sections, the magnitudes of both the CP asymmetry in the axino decay and the $n-\overline{n}$ oscillation time,
$\tau_{n\overline{n}}$, can reach the desired levels for $\mathcal{O}(0.1)$ values of the RPV coupling $\lambda''_{313}$. Additionally, a large $\tau_{n\overline{n}}$ requires supersymmetric partners with masses $\lesssim 1$ TeV. 
Such a light spectrum has been extensively sought by the LHC experiments, 
although usual searches for supersymmetry requiring large missing momentum are insensitive to our case where R-parity is violated. 
In fact, although the axino LSP is long-lived on detector scales, the heavier supersymmetric particles, if produced in $pp$ collisions at the LHC, would eventually decay to SM quarks through $\lambda''_{313}$ rather than into the axino whose couplings are suppressed by the large scale $f_a$. 
Hence, the LHC phenomenology is dictated by the nature of the next-to-LSP (NLSP): heavier particles undergo decay chains ending with the NLSP, which decays to SM quarks. In particular, among the particles involved in the processes we are interested in, a stop NLSP would simply decay through the  $\lambda''_{313}$ coupling as $\widetilde t_1\to \overline{b}\,\overline{d}$ (analogously
for a sbottom NSLP $\widetilde b_1\to \overline{t}\,\overline{d}$), while if the NSLP is a gaugino, such as the Wino, it would decay to three bodies via an off-shell squark, e.g.~$\widetilde W^0 \to \overline{t}\,\widetilde{t}_1 \to \overline{t}\, \overline{b}\,\overline{d}$. 
In the large coupling regime we are interested in, $\lambda''_{313}=\mathcal{O}(0.1)$, both the above decays have large enough rates to occur promptly at the $pp$ interaction point.
\begin{figure}[t]
\begin{center}
\includegraphics[width=.45\columnwidth]{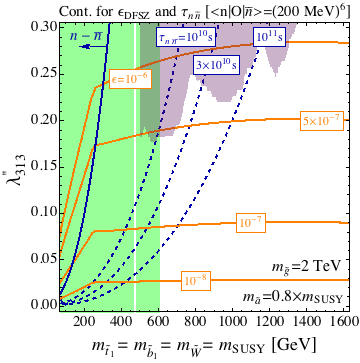}
\hspace{0.3cm}
\includegraphics[width=.45\columnwidth]{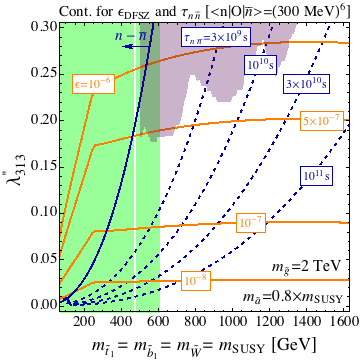}\\
\vspace{0.3cm}
\includegraphics[width=.45\columnwidth]{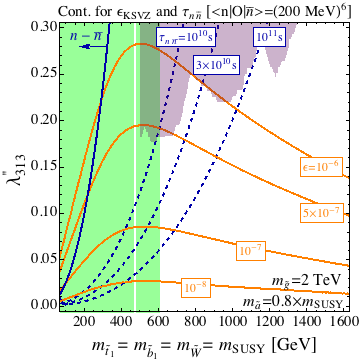}
\hspace{0.3cm}
\includegraphics[width=.45\columnwidth]{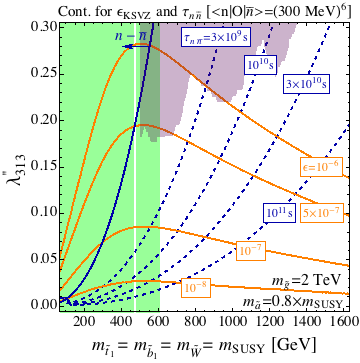}
\caption{\footnotesize Exclusion on the $n-\overline{n}$ oscillation time $\tau_{n\overline{n}}$ from \cite{nnbar-SK} (solid blue line) and contours for several values of 
 $\tau_{n\overline{n}}$ (dashed blue line) and the CP asymmetry $\epsilon$ (orange lines) induced by the axino in the DFSZ (first row) and KSVZ case  (second row), displayed in the plane of a common mass $m_{\rm SUSY}$ and $\lambda''_{313}$. 
Stop and sbottom mixing are taken as $\theta_{\widetilde t} =\pi/6$ and $\theta_{\widetilde b} =\pi/4$, and $\varphi_{\widetilde a}=1$ is the only non-vanishing phase.
The other relevant parameters are as indicated in the plots. The light-green area is excluded by the ATLAS four-jets search \cite{ATLAS:2017gsy}. The purple area is excluded by resonant stop production \cite{Monteux:2016gag}. 
\label{fig:plot1}}
\end{center}
\end{figure}

A recent search (based on the full data-set of the 2016 LHC run at $\sqrt{s}=13$ TeV)
 for pair-produced resonances each decaying into two jets (including b-jets) \cite{ATLAS:2017gsy} -- thus sensitive to direct production of stop pairs with the above RPV decay mode $\widetilde t_1\to \overline{b}\,\overline{d}$ -- excludes stop masses in the range $100~{\rm GeV}< m_{\widetilde t_1} < 470~{\rm GeV}$ and $480~{\rm GeV}< m_{\widetilde t_1} < 610~{\rm GeV}$ (the gap being due to what appears to be a slight statistical fluctuation).
Similarly, other recent searches based on events with large jet multiplicities \cite{ATLAS:2017wgj,Aaboud:2017hdf} can 
be interpreted in terms of production of gluinos decaying into a top and a RPV-decaying stop, resulting in a limit on the gluino mass up to $m_{\widetilde g}\lesssim 1.6$ TeV. The search of \cite{ATLAS:2017wgj} has been also interpreted to constrain the case of stop production with the stops decaying into lighter charginos and neutralinos, hence addressing in our case the possibility of a Wino NLSP with the above-mentioned three-body decay. 
This sets a limit on the stop mass up to 1.1 TeV, but no bound is placed for a stop-gaugino mass splitting smaller than $m_t$, and similarly
 the sensitivity is rapidly lost if the stop is lighter than about 600 GeV. 
Finally, large RPV couplings can induce resonant single squark production at sizable rates, e.g.~in our case $d\,b \to {\widetilde t_1}^*$. Based on this, several LHC searches with 8 TeV and early 13 TeV data have been employed in \cite{Monteux:2016gag} to obtain upper limits on the $\lambda''_{ijk}$ couplings as a function of the squark mass: in particular $\lambda''_{313}\lesssim 0.2$ for $ m_{\widetilde t_1} < 1~{\rm TeV}$.

\begin{figure}[t]
\begin{center}
\includegraphics[width=.45\columnwidth]{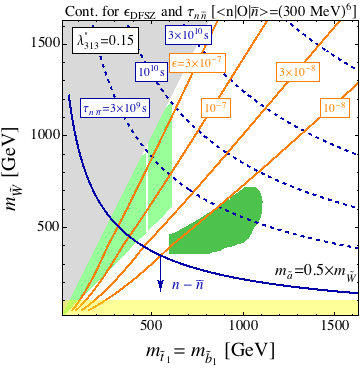}
\hspace{0.3cm}
\includegraphics[width=.45\columnwidth]{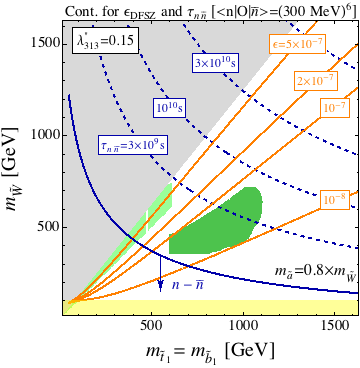}
\caption{\footnotesize Contours for  $\tau_{n\overline{n}}$ and $\epsilon$ in the DFSZ scenario in the
 $m_{\widetilde t_1}(=m_{\widetilde b_1})$--$m_{\widetilde W}$. The values of the parameters were taken as in Fig.~\ref{fig:plot1} unless otherwise indicated. 
The light-green area is excluded by the ATLAS four-jets search \cite{ATLAS:2017gsy}, the dark-green region by the multijets search \cite{ATLAS:2017wgj}.
The yellow region is excluded by LEP, while  in the gray-shaded area the axino is not the LSP.
\label{fig:plot2}}
\end{center}
\end{figure}

The impact of the these searches on our parameter space is depicted in Figs.~\ref{fig:plot1} and~\ref{fig:plot2}. 
In Fig.~\ref{fig:plot1}, we plot contours of the CP asymmetry in the decay of the DFSZ (first row) and KSVZ axino (second row) as a function of the coupling and a common mass $m_{\widetilde t_1}=m_{\widetilde b_1}=m_{\widetilde W}$,
 together with the prediction for the $n-\overline{n}$ oscillation time.  As we can see from Eq.~(\ref{eq:nnbar-num}), 
this observable strongly depends on the matrix element  $\langle  n| (udd)^2 |\overline n\rangle$ whose value
at present can be only estimated to be of the order of $\Lambda_{\rm QCD}^6$. In order to take into account this large uncertainty affecting any prediction 
for $n-\overline{n}$ oscillations, we chose to show two `extremal' values, $\langle  n| (udd)^2 |\overline n\rangle = (200~{\rm MeV})^6$ (left plots)
and $(300~{\rm MeV})^6$ (right plots), for which $\tau_{n\overline{n}}$ approximately spans one order of magnitude.
The mass of the axino LSP is taken $m_{\widetilde a} = 0.8\times m_{\rm SUSY}$, and large stop and sbottom mixing (respectively $\theta_{\widetilde t} =\pi/6$ and $\theta_{\widetilde b} =\pi/4$) as well as $\varphi_{\widetilde a}=1$ for the axino phase are assumed.
The areas excluded by the four-jets search \cite{ATLAS:2017gsy} and resonant stop production \cite{Monteux:2016gag}  are shown in light green and purple respectively, while multijets searches  \cite{ATLAS:2017wgj,Aaboud:2017hdf} are evaded for the heavy gluino mass $m_{\widetilde g}=2$ TeV that we chose.
Although these LHC constraints are largely affecting our parameter space, we can see that they still leave room to values of the CP asymmetry,
$\epsilon \approx 10^{-7}- 10^{-6}$, able to induce a successful baryogenesis (as discussed in Sections \ref{sec:DFSZ} and \ref{sec:KSVZ}), with, at the same time,
$n-\overline{n}$ oscillation times at the level of $\tau_{n\overline{n}} \approx 10^9-10^{10}~\sec$. 
Furthermore, the bound from the four-jets search (light-green region) can be relaxed and eventually evaded taking $m_{\widetilde W} < m_{\widetilde t_1}$,
as the stop will then preferably decay to neutralino or chargino. As we mentioned above, searches as in \cite{ATLAS:2017wgj} can be in turn sensitive to the RPV decays of the Wino, but only for rather heavy MSSM particles and $m_{\widetilde t_1} -m_{\widetilde W}  > m_t$. This is better illustrated by Fig.~\ref{fig:plot2},
where we set  $\lambda''_{313}= 0.15$ and we plot our observables in the plane $m_{\widetilde t_1}$--$m_{\widetilde W}$,
for two different values of the axino mass $m_{\widetilde a} = 0.5\times m_{\widetilde W}$ (left panel) and $m_{\widetilde a} = 0.8\times m_{\widetilde W}$ (right panel). Again the exclusion from the four-jets search is shown in light green, while the area to which the multijets search \cite{ATLAS:2017wgj} 
is sensitive is dark green. Additionally, in the gray-shaded area the axino is not the LSP and the yellow band represents the LEP bound on the mass of charginos. 

To summarize the above discussion, our scenario, despite the LHC constraints, can still achieve a large CP asymmetry (thus triggering LSP baryogenesis) and large $n-\overline{n}$ oscillation rates at the same time, provided that 
\begin{itemize}
\item the RPV coupling is $\mathcal{O}(0.1)$; 
\item stop and sbottom left-right mixing is large;
\item  an $\mathcal{O}(1)$ phase is present without inducing further constraints from CP-violating observables (a good example being the axino mass phase);
\item  the masses of the relevant supersymmetric particles, in particular ${\widetilde t_1}$, ${\widetilde b_1}$ and ${\widetilde W}$ are $< 1$ TeV, thus
at the level of the current sensitivity of searches for RPV supersymmetry at the LHC.
\end{itemize}

\section{Conclusions}
\label{sec:conclusions}
We have presented a scenario that consistently accounts for some formidable shortcomings of the Standard Model: the baryon asymmetry of the Universe, the observed amount of Dark Matter, the strong CP problem, as well as the hierarchy problem.
The baryon asymmetry can be successfully produced by the decay of the axino LSP to SM quarks through RPV interactions,
while DM and the strong CP problem are accounted for by the axion. We have shown that this is the case for both the DFSZ and the KSVZ axion models, although axion DM prefers the DFSZ scenario. 
In fact, in the DFSZ case, the requirement that the axino decays do not spoil BBN, i.e.~$T_{D}>T_{\rm BBN}$, and the more stringent one of having a sizable baryon asymmetry, $T_D \gtrsim \mathcal{O}(0.1-1)$ GeV 
are remarkably fulfilled for values of the PQ scale $f_a$ compatible with the observed DM abundance through the vacuum realignment mechanism of the axion field, although only for rather large values of the misalignment angle. 

In our scenario, the LSP baryogenesis mechanism can be realized for a supersymmetric spectrum at the TeV scale. As we have shown, the same BNV interactions required by baryogenesis can then induce $\Delta B =2$ processes such as neutron-antineutron at rates potentially observable by next-generation experiments. At the same time, large production rates of colored superpartners, in particular stops and gluinos, are possible at the LHC. 
This opens up the exciting possibility of testing our scenario in a number of experiments at very different energies: 
$n-\overline{n}$, LHC, as well as axion search experiments.


\acknowledgments
We thank G.~Ferretti and C.~Petersson for useful discussions. LC is grateful to KIAS -- where this work was initiated -- for the kind hospitality and financial support.
CSS acknowledges  the support from the Korea Ministry of Education, Science and Technology, Gyeongsangbuk-Do and Pohang City for Independent Junior Research Groups at the Asia Pacific Center for Theoretical Physics. CSS is also supported by the Basic Science Research Program through the NRF Grant (No. 2017R1D1A1B04032316).

\newpage


\end{document}